\documentclass[12pt]{article}
\usepackage[dvips]{graphicx}
\usepackage{amsmath}

\title{Regularization for zeta functions II}
\author{}

\begin{document}
{\pagestyle{empty}
\rightline{May 2012}
\rightline{~~~}
\vskip 1cm
\centerline{\large \bf Regularization for zeta functions with physical applications II}
\vskip 1cm
\centerline{{Minoru Fujimoto\footnote{E-mail address: 
             seikakagaku@kcn.jp}} and
            {Kunihiko Uehara\footnote{E-mail address: 
             uehara@tezukayama-u.ac.jp}}}
\vskip 1cm
\centerline{\it ${}^1$Seika Science Research Laboratory,
Seika-cho, Kyoto 619-0237, Japan}
\centerline{\it ${}^2$Department of Physics, Tezukayama University,
Nara 631-8501, Japan}
\vskip 2cm

\centerline{\bf Abstract}
\vskip 0.2in
  We have proposed a regularization technique and applied it to 
the Euler product of zeta functions in the part one.
In this paper that is the second part of the trilogy, we give another evidence 
to demonstrate the Riemann hypotheses by using the approximate functional equation. 
Some other results on the critical line are presented using the relations 
between the Euler product and the deformed summation representations in the critical strip. 
The relations between the prime numbers and the zeros of the Riemann zeta functions are also referred. 
In part three, we will focus on physical applications using these outcomes.

\vskip 0.4cm\noindent
PACS number(s): 02.30.-f, 02.30.Gp, 05.40.-a

\hfil
\vfill
\newpage}
\setcounter{equation}{0}
\addtocounter{section}{0}
\section{Introduction}
\hspace{\parindent}
  In the situation that the regularizations by the zeta function have been 
successful with some physical applications, we proposed a regularization 
technique\cite{Fujimoto} applicable to the Euler product representation and gave an 
evidence for the Riemann hypotheses by using this technique in the part one.
In this part, we now focus on the zeros of the Riemann zeta function 
and the surrounding properties of the zeros including other evidences 
to demonstrate the Riemann hypotheses.
The Euler product representation, which played an essential role 
in the first part, will be interpreted in terms of the summation representation 
on the critical line $\Re{z}=\frac{1}{2}$.

  The definition of the Riemann zeta function is 
\begin{equation}
  \zeta(z)=\lim_{n\rightarrow\infty}\zeta_n(z)
          =\lim_{n\rightarrow\infty}\sum_{k=1}^n\frac{1}{k^z}
          =\prod_{k=1}^\infty\left(1-\frac{1}{{p_k}^z}\right)^{-1}
\label{e101}
\end{equation}
for $\Re z>1$, where the right hand side is the Euler product representation and 
$p_k$ is the $k$-th prime number. 
Hereafter we adopt a notation $\hat{\zeta}(z)$ for $\Re z>0$ such as
\begin{equation}
  \hat\zeta(z)=\lim_{n\rightarrow\infty}\hat{\zeta}_n(z)
              =\lim_{n\rightarrow\infty}
               \frac{1}{1-2^{1-z}}\sum_{k=1}^n\frac{(-1)^{k-1}}{k^z},
\label{e102}
\end{equation}
which is well regularized even in the critical strip $0<\Re z<1$. 

  Considering the approximate expansion formula for the Riemann zeta function, 
we propose an evidence for an elegant proof of the Riemann hypothesis 
in section 2. And in \S 3 we show surrounding properties of the zeros 
for the Riemann zeta function by deforming the Euler product representation
to the summation form on the critical line. 
We study the relation between the primes and the zeros of the zeta function in connection with 
the Sato-Tate conjecture in section 4, and 
we will discuss the equations for these primes and zeros in \S 5.

\hskip 5mm
\section{The expansion formula and the Riemann hyposesis}
\hspace{\parindent}

The Euler-Maclaurin sum formula is given by
\begin{equation}
  \sum_{n=M}^Nf(n)=\int_M^Nf(x)dx+\frac{1}{2}(f(M)+f(N))
                  +\sum_{j=1}^k\frac{B_{2j}}{(2j)!}
                   \left[f^{(2j-1)}(x)\right]_M^N
                  +R_{2k},
\label{e201}
\end{equation}
where $B_{2j}$ is the $2j$-th Bernoulli number and the remainder term: 
\begin{eqnarray}
  R_{2k}&=&\frac{1}{(2k+1)!}\int_M^N\bar{B}_{2k+1}(x)f^{(2k+1)}(x)dx,\nonumber\\
  \bar{B}_{2k+1}(x)&=&B_{2k+1}(x-[x]).
\label{e202}
\end{eqnarray}

We parametrize a complex variable $z$ by two real variable such as 
$\displaystyle{z=s(\frac{1}{2}+it)}$ as same as that in the first part.
Using the Euler-Maclaurin sum formula on the assumption of $\displaystyle\frac{s}{2}=\Re z>-2k$, 
we get the relation 
\begin{eqnarray}
  \sum_{n=M}^\infty\frac{1}{n^z}
    &=&\int_M^\infty x^{-z}dx+\frac{1}{2}M^{-z}
       +\int_M^\infty\bar{B}_1(x)(-z)x^{-z-1}dx\nonumber\\
    &=&\frac{M^{1-z}}{z-1}+\frac{1}{2}M^{-z}
       +\sum_{j=1}^k\frac{B_{2j}}{(2j)!}M^{1-z-2j}\prod_{l=0}^{2j-2}(z+l)
       +R_{2k},
\label{e203}
\end{eqnarray}
where 
\begin{equation}
  R_{2k}=\frac{-z(z+1)\cdots(z+2k)}{(2k+1)!}
         \int_M^\infty\bar{B}_{2k+1}(x)x^{-z-2k-1}dx.
\label{e204}
\end{equation}
As is well known we can go forward to the expansion formula 
in the case of $M\ge 2$, 
\begin{equation}
  \hat{\zeta}(z)=\sum_{n=1}^{M-1}\frac{1}{n^z}
          +\frac{M^{1-z}}{z-1}+\frac{1}{2}M^{-z}
          +\sum_{j=1}^k\frac{B_{2j}}{(2j)!}M^{1-z-2j}\prod_{l=0}^{2j-2}(z+l)
          +R_{2k}.
\label{e205}
\end{equation}
The remainder term $R_{2k}$ can be estimated as follows:
\begin{eqnarray}
  |R_{2k}|&\le&\frac{\pi^2}{3}\left|\frac{z+2k+1}{\frac{1}{2}+2k+1}\right|
               \left|\frac{z(z+1)\cdots(z+2k)}{(2\pi)^{2k+2}}\right|
               M^{-\frac{1}{2}-2k-1}\nonumber\\
          &\le&C(k)\sqrt{M}\left(\frac{t}{2\pi M}\right)^{2k+2},
\label{e206}
\end{eqnarray}
where we put $s=1$ and $C(k)$ is constant only depending on $k$. This tells us that 
it is necessary for the remainder term  to be 
$M>\displaystyle\frac{t}{2\pi}$ to converge.

Taking account of the Euler-Maclaurin sum formula, we can put 
the regularized zeta function as
\begin{equation}
  \lim_{n\rightarrow\infty}\hat{\zeta}_n(z)
  =\lim_{n\rightarrow\infty}\left\{\zeta_n(z)-\frac{n^{1-z}}{1-z}\right\}
\label{e207}
\end{equation}
and a zero in the critical strip is the solution to the equation 
\begin{equation}
  \hat{\zeta}(z)=\lim_{n\rightarrow\infty}\hat{\zeta}_n(z)=0.
\label{e208}
\end{equation}
These equations (\ref{e207}) and (\ref{e208}) are identical to the equation (A4) 
in the first part. 
As stated in Appendix A in the first part, Eq.(\ref{e102}) can be derived 
by way of the regularization method developed in the part one, 
which means that we can reach here besides using the Euler-Maclaurin sum formula.
As $(1-\rho)$ is also the solution when $\rho$ is the solution 
of Eq.(\ref{e208}), a solution of the equation
\begin{equation}
  \hat{\zeta}(1-z)=0
\label{e209}
\end{equation}
is also a zero.
Now we transform Eq.(\ref{e207}) to
\begin{equation}
  \lim_{n\rightarrow\infty}\{(1-z)\zeta_n(z)-n^{1-z}\}=0
\label{e210}
\end{equation}
and substituting $(1-z)$ for $z$, we get
\begin{equation}
  \lim_{n\rightarrow\infty}\{z\zeta_n(1-z)-n^z\}=0.
\label{e211}
\end{equation}
Combining these equations (\ref{e210}) and (\ref{e211}), we get
\begin{equation}
  \lim_{n\rightarrow\infty}\{z(1-z)\zeta_n(z)\zeta_n(1-z)-n\}=0,
\label{e212}
\end{equation}
namely,
\begin{equation}
  z^2-z+\lim_{n\rightarrow\infty}\frac{n}{\zeta_n(z)\zeta_n(1-z)}=0.
\label{e213}
\end{equation}
The solution $\rho_n$ of the equation $\hat{\zeta}_n(z)=0$ 
is satisfied the relation
\begin{equation}
  \rho_n=\frac{1}{2}
         \pm i\sqrt{
                    \frac{n}{\zeta_n(\rho_n)\zeta_n(1-\rho_n)}}.
\label{e2135}
\end{equation}

On the other hand, the approximate functional equation by Hardy and Littlewood\cite{Hardy}, 
which leads to the Riemann-Siegel formula, is given by
\begin{equation}
  \hat{\zeta}(z)=\sum_{n\le x}\frac{1}{n^z}
          +\hat{H}(z)\sum_{n\le y}\frac{1}{n^{1-z}}
          +O(x^{-s/2})+O(|t|^{1/2-s/2}y^{s/2-1}),
\label{e214}
\end{equation}
where $0\le s/2(=\Re z)\le 1, x\ge 1, y\ge 1, 2\pi xy=|t|$ and 
$\hat{H}(z)$ is given by
\begin{equation}
  \hat{H}(z)=2\Gamma(1-z)(2\pi)^{z-1}\sin\frac{\pi z}{2}.
\label{e215}
\end{equation}
For $s>2$, the relation
\begin{equation}
  \zeta(z)=H(z)\sum_{n=1}^\infty\frac{1}{n^{1-z}}
          =H(z)\zeta(1-z)
\label{e215}
\end{equation}
is satisfied and we write down $\hat{\zeta}(1-z)$ for $0<\Re z<1$ as, 
\begin{equation}
  \hat{\zeta}(1-z)=\frac{1}{1-2^z}\sum_{n=1}^\infty\frac{(-1)^{n-1}}{n^{1-z}}.
\label{e216}
\end{equation}
Then the relation
\begin{equation}
  \hat{\zeta}(z)=\hat{H}(z)\hat{\zeta}(1-z)
\label{e217}
\end{equation}
is satisfied for $\Re z>0$ and substituting $(1-z)$ for $z$, 
we conclude Eq.(\ref{e217}) is satisfied for all $z$.

  We set $n\le x, n\le y$ and $|t|=2\pi xy\ge 2\pi n^2>2\pi n$, then 
\begin{equation}
  \hat{\zeta}(z)=\zeta_n(z)+\hat{H}(z)\zeta_n(1-z)+R_n(z),
\label{e218}
\end{equation}
where the remainder term $R_n(z)$:
\begin{eqnarray}
  R_n(z)&=&O(n^{-s/2})+O(|t|^{1/2-s/2}n^{s/2-1})\nonumber\\
        &=&O(n^{-s/2})+o((2\pi n)^{1/2-s/2}n^{s/2-1})
\label{e219}
\end{eqnarray}
can be ignored by taking the limit of $n\rightarrow\infty$, namely
\begin{equation}
  \lim_{n\rightarrow\infty}R_n(z)=
    \lim_{n\rightarrow\infty}\{O(n^{-s/2})+o(n^{-1/2})\}
    =0.
\label{e220}
\end{equation}

We put the relations at zeros 
\begin{equation}
  \zeta_n(\rho_n)=\frac{n^{1-\rho_n}}{1-\rho_n},\quad
  \zeta_n(1-\rho_n)=\frac{n^{\rho_n}}{\rho_n}
\label{e221}
\end{equation}
into Eq.(\ref{e218}) together with $\hat{\zeta}_n(\rho_n)=0$, we can write 
\begin{eqnarray}
  \frac{n^{1-\rho_n}}{1-\rho_n}
  +\hat{H}(\rho_n)\frac{n^{\rho_n}}{\rho_n}+R_n(\rho_n)&=&0,\nonumber\\
  \frac{n^{1-2\rho_n}}{1-\rho_n}
  +\hat{H}(\rho_n)\frac{1}{\rho_n}+R_n(\rho_n)\frac{1}{n^{\rho_n}}&=&0.
\label{e222}
\end{eqnarray}
Taking the limit of $n\rightarrow\infty$, we get the relations
\begin{equation}
  \lim_{n\rightarrow\infty}\left|\frac{n^{1-2\rho_n}}{1-\rho_n}\right|
  =-\lim_{n\rightarrow\infty}\left|\frac{\hat{H}(\rho_n)}{\rho_n}\right|
  =-\left|\frac{\hat{H}(\rho)}{\rho}\right|.
\label{e223}
\end{equation}
The right hand side of Eq.(\ref{e223}) is finite, so the numerator of 
the left hand side $\displaystyle\lim_{n\rightarrow\infty}|n^{1-2\rho_n}|$ must be finite. 
This means that the real part of $(1-2\rho_n)$ must converge to $0$ 
in the limit of $n\rightarrow\infty$ when the real part of $(1-2\rho_n)$ is positive.
When $\Re(1-2\rho_n)<0$, we rewrite the left hand side of Eq.(\ref{e223}) like 
\begin{equation}
  \lim_{n\rightarrow\infty}\frac{n^{1-2(1-\rho_n)}}{1-(1-\rho_n)}
  =\lim_{n\rightarrow\infty}\frac{n^{2\rho_n-1}}{\rho_n},
\label{e2235}
\end{equation}
so we get the same goal as $\displaystyle{\lim_{n\rightarrow\infty}}\Re(2\rho_n-1)=0$.
After all, the Riemann hyposesis is satisfied including the trivial case 
$\Re(1-2\rho_n)=0$, 
\begin{equation}
  \rho=\lim_{n\rightarrow\infty}\rho_n
      =\lim_{n\rightarrow\infty}\left(\frac{1}{2}+it_n\right)
      =\frac{1}{2}+i\lambda,
\label{e224}
\end{equation}
where $\lambda$ is real but $t_n$ is not necessarily a real number.

Now we think about the values of $t_n$, which converge to a positive $\lambda$ 
in the limit of $n\rightarrow\infty$ and put it into Eq.(\ref{e2135})
\begin{eqnarray}
  \rho_n&=&\frac{1}{2}
         +i\sqrt{\frac{n}{\zeta_n(\rho_n)\zeta_n(1-\rho_n)}}\nonumber\\
        &=&\frac{1}{2}
         +i\sqrt{\frac{n}{\zeta_n(\frac{1}{2}+it_n)\zeta_n(\frac{1}{2}-it_n)}}.
\label{e225}
\end{eqnarray}
Thus we write the solution to the equation $\hat{\zeta}(z)=0$
\begin{eqnarray}
  \rho&=&\lim_{n\rightarrow\infty}\left\{\frac{1}{2}
         +i\sqrt{\frac{n}{\zeta_n(\rho_n)\zeta_n(1-\rho_n)}}\right\}\nonumber\\
  &\mbox{\it i.e.}&
\label{e226}\\
  \lambda&=&\lim_{n\rightarrow\infty}
    \sqrt{\frac{n}{\zeta_n(\frac{1}{2}+it_n)\zeta_n(\frac{1}{2}-it_n)}}\nonumber.
\end{eqnarray}
Using the n-th order relation of Eq.(\ref{e217})
\begin{equation}
  \zeta_n(\frac{1}{2}-it_n)
    =H(\frac{1}{2}-it_n)\zeta_n(\frac{1}{2}+it_n),
\label{e227}
\end{equation}
we get the relation 
\begin{eqnarray}
  t_n&=&\sqrt{
         \frac{n}{
          \zeta_n(\frac{1}{2}+it_n)
          H(\frac{1}{2}-it_n)
          \zeta_n(\frac{1}{2}+it_n)
         }
        }\nonumber\\
     &=&\frac{1}{\zeta_n(\frac{1}{2}+it_n)}
        \sqrt{\frac{n}{H(\frac{1}{2}-it_n)}}.
\label{e228}
\end{eqnarray}
This form will be utilize to calculate zeros of the Riemann zeta function 
by way of the limit of $n\rightarrow\infty$.

\vskip 5mm
\section{The Euler product and a summation representation}
\hspace{\parindent}
We write down the Euler product representation for the standard form 
described as same as the equation (25) in the first part 
\begin{equation}
  \frac{1}{\left|\zeta_n(s(\frac{1}{2}+it))\right|^2}
  =f_n(s,t)
  =\prod_{k=1}^n\left(1-\frac{2}{p_k^{s/2}}\cos(st\log p_k)
   +\frac{1}{{p_k}^s}\right).
\label{e301}
\end{equation}
Here $f_n(s,t)$ and $\log f_n(s,t)$ will diverge at the same time in $n\rightarrow\infty$, 
because $f_n(s,t)$ is positive.
As all non-trivial zeros of the Riemann zeta function is expected on $s=1$, 
we study the following relation for $s\ge 1$
\begin{eqnarray}
  \log f_n(s,t)
  &=&\sum_{k=1}^n\log\left(1-\frac{2}{{p_k}^{s/2}}\cos(st\log p_k)
   +\frac{1}{{p_k}^s}\right)\nonumber\\
  &=&\sum_{k=1}^n\log\left\{1-\left(\frac{2}{{p_k}^{s/2}}\cos(st\log p_k)
   -\frac{1}{{p_k}^s}\right)\right\}\nonumber\\
  &=&-\sum_{k=1}^n\sum_{m=1}^\infty\frac{1}{m}\left(\frac{2}{{p_k}^{s/2}}\cos(st\log p_k)
   -\frac{1}{{p_k}^s}\right)^m\nonumber\\
  &=&-\sum_{k=1}^n\left(\frac{2}{{p_k}^{s/2}}\cos(st\log p_k)-\frac{1}{{p_k}^s}\right)
   -\frac{1}{2}\sum_{k=1}^n\left(\frac{2}{{p_k}^{s/2}}\cos(st\log p_k)
    -\frac{1}{{p_k}^s}\right)^2\nonumber\\
   &&-\frac{1}{3}\sum_{k=1}^n\left(\frac{2}{{p_k}^{s/2}}\cos(st\log p_k)
    -\frac{1}{{p_k}^s}\right)^3
     +\text{(finite terms for $n\rightarrow\infty$)}.
\label{e302}
\end{eqnarray}

We must regularize Eq.(\ref{e302}) in order to apply it even in the case $s=1$.
We try to regularize Eq.(\ref{e302}) by way of dividing an appropriate factor
$\displaystyle\sum_{k=1}^n\frac{1}{p_k}$,
which leaves a leading divergence divergent and makes a non-leading divergence convergent.
In fact, we divide Eq.(\ref{e302}) by
\begin{equation}
  \prod_{k=1}^n\left(1+\frac{1}{p_k}\right),
\label{e303}
\end{equation}
which we also adopted in the equation (26) of the first part.\cite{Fujimoto} 
Thus we study the divergence in the form of 
\begin{equation}
  \frac{\displaystyle \prod_{k=1}^n\left(1-\frac{2}{p_k^{s/2}}\cos(st\log p_k)+\frac{1}{{p_k}^s}\right)}
       {\displaystyle \prod_{k=1}^n\left(1+\frac{1}{p_k}\right)},
\label{e304}
\end{equation}
which corresponds in the summation form as
\begin{equation}
  \frac{\displaystyle -2\sum_{k=1}^n\frac{\cos(t\log p_k)}{\sqrt{p_k}}
        \left(\frac{\cos(t\log p_k)}{\sqrt{p_k}}+1\right)+\sum_{k=1}^n\frac{1}{p_k}}
       {\displaystyle \sum_{k=1}^n\frac{1}{p_k}},
\label{e305}
\end{equation}
where we set $s=1$.
The leading divergent term of Eq.(\ref{e305}) in $n\rightarrow\infty$ is 
\begin{equation}
  \sum_{k=1}^n\frac{\cos(t\log p_k)}{\sqrt{p_k}}.
\label{e306}
\end{equation}

The form of divisor 
$\displaystyle{\prod_{k=1}^n\left(1+\frac{1}{p_k}\right)}$ 
means that using the Mertens' theorem
\begin{equation}
  \prod_{k=1}^m\left(1-\frac{1}{p_k}\right)
   =\frac{e^{-\gamma}}{\log p_m}\left(1+O\left(\frac{1}{\sqrt{p_m}}\right)\right)
\label{e307}
\end{equation}
and the Euler's $\zeta(2)$
\begin{equation}
  \prod_{k=1}^\infty\left(1-\frac{1}{{p_k}^2}\right)=\frac{6}{\pi^2},
\label{e308}
\end{equation}
we get 
\begin{eqnarray}
  C\prod_{k=1}^m\left(1+\frac{1}{p_k}\right)
  &=&\prod_{k=1}^m\frac{\left(1-\frac{1}{{p_k}^2}\right)}{\left(1-\frac{1}{p_k}\right)}
     \prod_{k=m+1}^\infty\left(1-\frac{1}{{p_k}^2}\right)\nonumber\\
  &=&\frac{6e^\gamma}{\pi^2}\log p_m\left(1+O\left(\frac{1}{\sqrt{p_m}}\right)\right),
\label{e3091}\\
  \log\prod_{k=1}^m\left(1+\frac{1}{p_k}\right)
  &=&\sum_{k=1}^m\log\left(1+\frac{1}{p_k}\right)\nonumber\\
  &=&\sum_{k=1}^m\frac{1}{p_k}
    -\frac{1}{2}\left(\sum_{k=1}^m\frac{1}{p_k}\right)^2+\cdots\nonumber\\
  &\simeq&\log\log p_m,
\label{e309}
\end{eqnarray}
where
\begin{equation}
  \frac{6e^\gamma}{\pi^2}\le C\le e^\gamma.
\label{e310}
\end{equation}

\begin{figure}[!b]
  \begin{center}
    \includegraphics[width=15cm,height=7cm,clip]{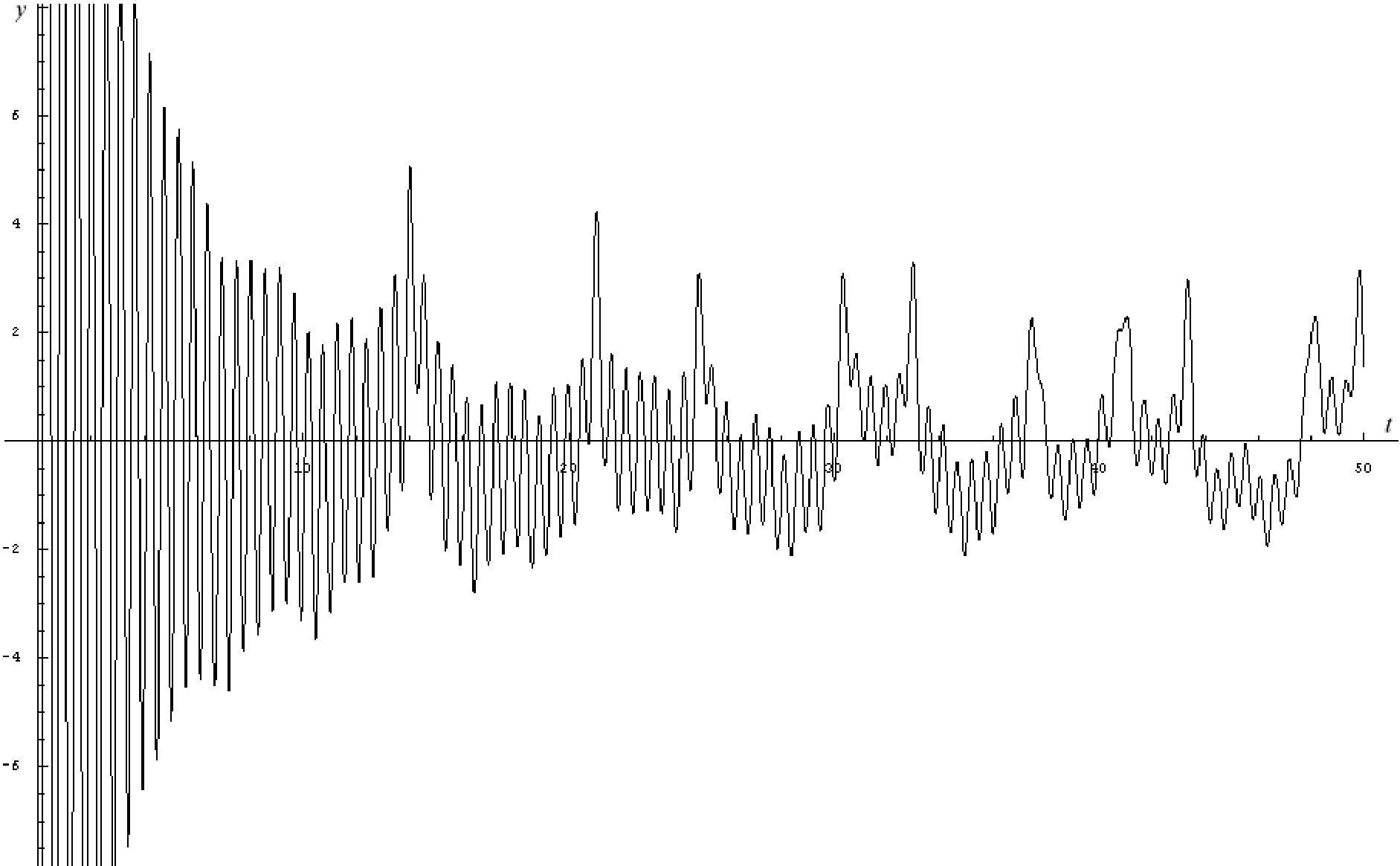}
  \end{center}
  \caption{\footnotesize The graph of $y_{n,\alpha}$ for $n=10^4$ 
           and $\alpha=\frac{1}{2}$.}
\end{figure}

\begin{figure}[!b]
  \begin{center}
    \includegraphics[width=15cm,height=7cm,clip]{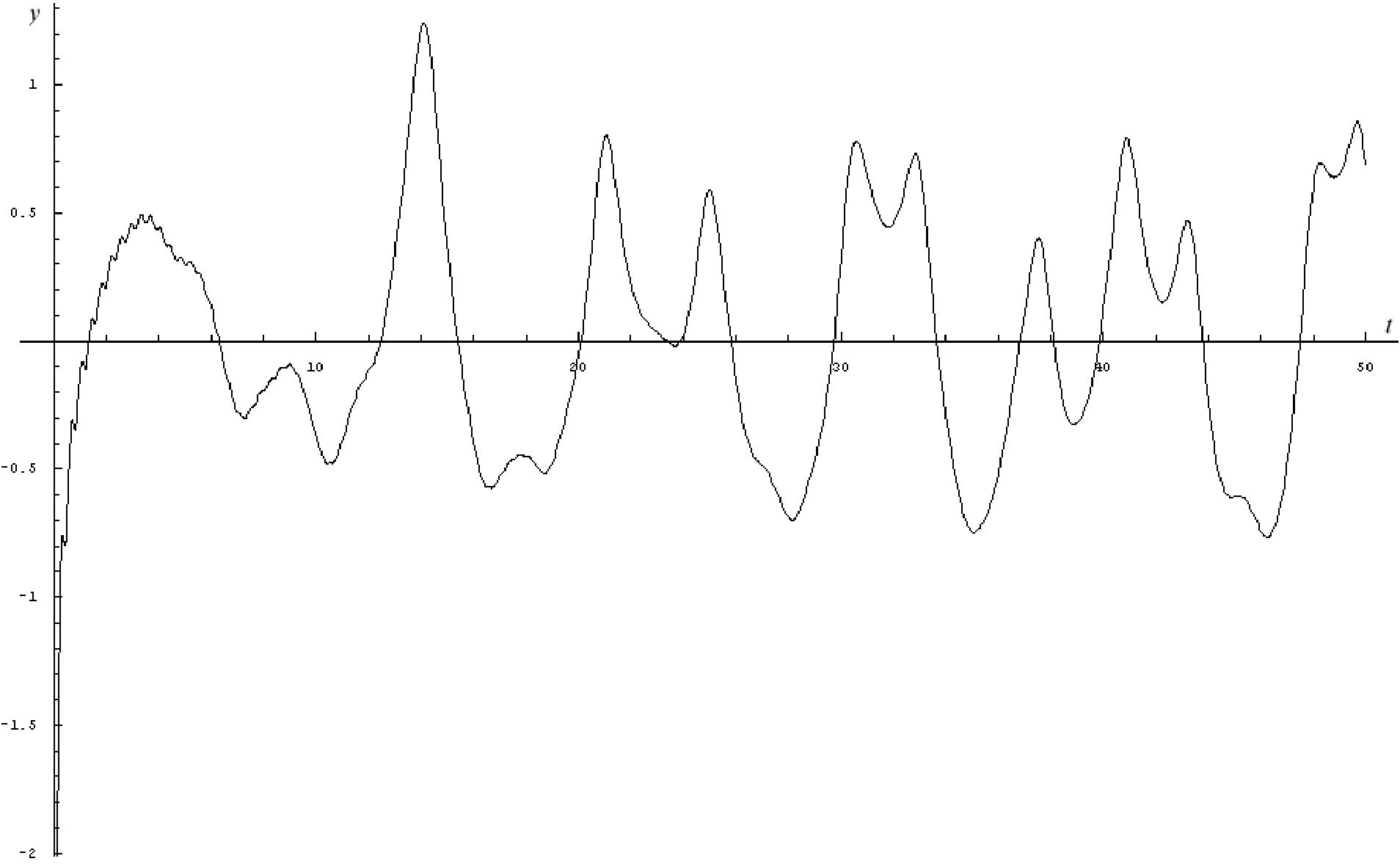}
  \end{center}
  \caption{\footnotesize The graph of $y_{n,\alpha}$ for $n=10^6$ 
           and $\alpha=1$.}
\end{figure}

The Euler product representation for $n\rightarrow\infty$ is only valid for 
$s\ge 2$, and we restrict our interest for $t>0$.
The zeros of the Riemann zeta function make Eq.(\ref{e301}) divergent, 
which means that products are multiplied maximally in the right hand side.
Each term in Eq.(\ref{e301}) is maximized when $\cos(st\log p_k)=-1$, namely, 
$\displaystyle st=\frac{(2\ell-1)\pi}{\log p_k}\quad\mbox{($\ell=$natural number)}$.
We give graphs for the superposition of cosine functions, 
which indicate the solution of $\cos(t\log p_k)=-1$ as local maximum values,
\begin{equation}
  y_{n,\alpha}(t)=-\sum_{k=1}^n\frac{\cos(t\log p_k)}{{p_k}^\alpha}.
\label{e311}
\end{equation}
\noindent
The graph of $y_{n,\alpha}(t)$ for $\alpha=1/2$ is printed as Figure 1, and 
judging from the graph of $\alpha=1$ (Figure 2), 
the denominator $p_k$ seems to be well-matched to cancel the notches come 
from the superposition of cosine functions. 
Figure 1 is also such an example of notches. 
Figure 2 has the positive maximal values that correspond to the non-trivial zeros of the zeta function 
except the one appeared in $t<6$. 
Thus zeros of the Euler product representation in Eq.(\ref{e301}) preserve 
the value even in the form of the summation in Eq.(\ref{e311}). 
The terms to regularize the divergence will be discussed, 
which is essential to the order on the critical line and seems to be closely related to the von Mangoldt function, 
in a separate paper. 

  On the other hand, the sum over the zeros of the zeta function for a certain prime $p$ 
\begin{equation}
  -2\sum_{j=1}^n\sqrt{p}\ {\cos(\lambda_j\log p)}
\label{e312}
\end{equation}
\noindent
leads us a graph which indicates locations of the prime numbers.\cite{Conrey} 

\vskip 5mm
\section{Nature of the prime numbers}
\hspace{\parindent}
Here we can show that the Riemann hypothesis holds for the $L$-function 
by using the approximate functional equation for the Dirichlet's $L$-function 
as well as that by using the regularization for the Euler product as stated in part one.
We listed the condition which leads a verification along these lines for the Riemann hypotheses as 
\begin{enumerate} 
  \item the existence of the Euler product representation, 
  \item the prime number theorem $\displaystyle\pi(x)\simeq\frac{x}{\log x}$ is satisfied,
  \item the approximate functional equation of the Dirichlet's $L$-function is 
satisfied.
\end{enumerate}
Exclusive uses of the Euler-Maclaurin expansion for the zeta function, which is actually 
an asymptotic expansion, have prevented the Riemann hypothesis from being demonstrated. 

According to the conclusion of the first part, the Riemann hypothesis for 
the Ramanujan's zeta function or another zeta function is realized 
because each function has the Euler product representation.
The Ramanujan's conjecture for the Euler product corresponds the cosine term 
of the standard form for the Riemann zeta function, 
so it will hold because $|\cos\theta|$ is less than one due to 
the independence of $\log p_k$'s.

About the zeta functions, which have no non-trivial zero besides zeros 
of the Riemann hypotheses, we parametrize them to the standard form.
In this case, the product of the zeros $\lambda_j$ of 
the Riemann zeta function and $\log p_k$, the logarithm of the primes $p_k$ 
has a similar structure to $\theta$ of the Sato-Tate conjecture or 
the Sato-Tate theorem for the zeta function associated with 
the elliptical function proved by Richard Taylor.
Moreover, in spite that the $\lambda_j$'s obey the uniform distribution 
to modulus one\cite{Elliot}, 
we claim that the response of $j$-direction increase($j=1,2,\cdots,\infty$) 
for $\lambda_j$ yields the similar distribution of the Sato-Tate 
conjecture,\cite{Sato} once we take $\lambda_j\log p_k$ to modulus $2\pi$.
The Sato-Tate conjecture claims that the response of 
$k$-direction increase($k=1,2,\cdots,\infty$) 
for $p_k$ yields the distribution of 
\begin{equation}
  \frac{2}{\pi}\int_\alpha^\beta\sin^2\theta d\theta,
\label{e600}
\end{equation}
where $0\le\alpha\le\theta\le\beta\le\pi$. 
On the other hand, once we put $2\theta=\lambda_j\log p_k$, 
we may claim that the response of $j$-direction increase of $\lambda_j$ yields 
the distribution of the Wigner's semi-circle law, which is related by 
regarding $\cos\theta$ of Eq.(\ref{e600}) as a single variable.

\begin{figure}[!t]
  \begin{center}
    \includegraphics[width=17cm,height=7cm,clip]{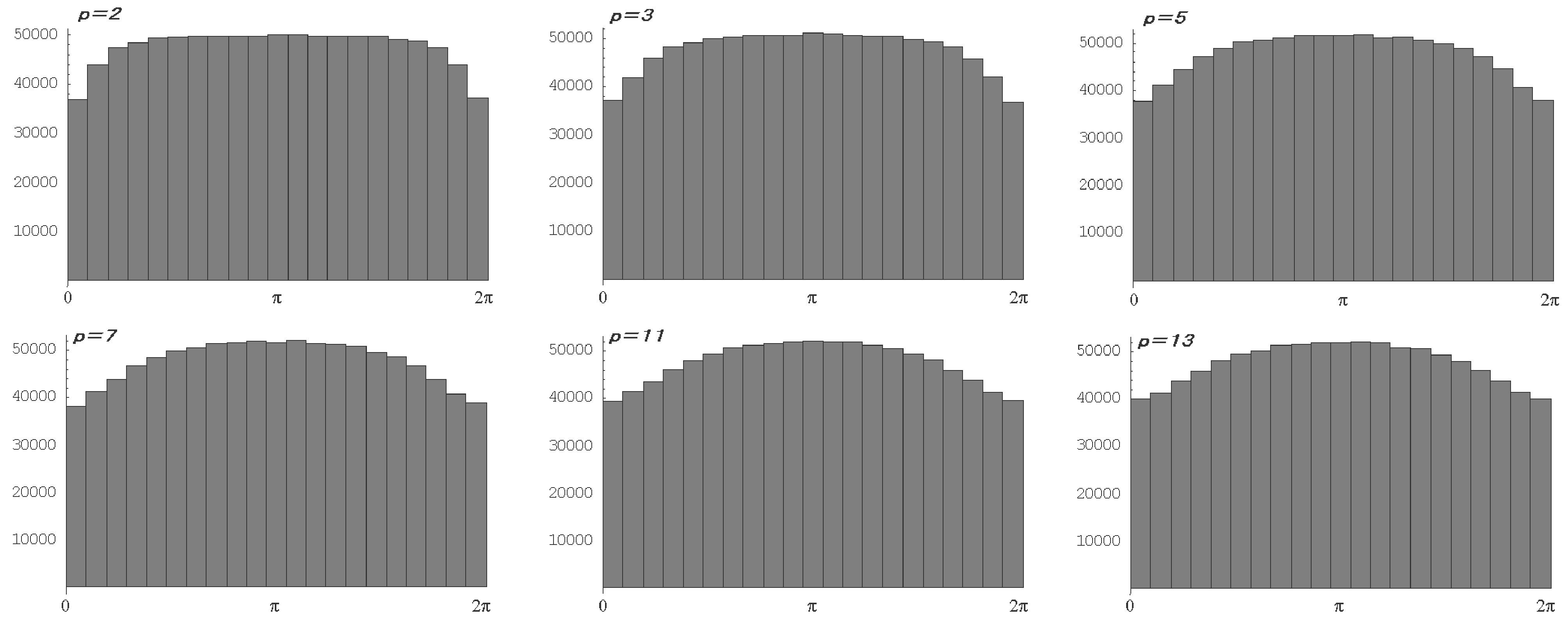}
  \end{center}
  \caption{\footnotesize The histograms(divided $21$-th) of distributions for 
           $p_k=2,3,5,7,11$ and $13$ beginning with $j=10^6$ up to $2\times 10^6$.}
\end{figure}

\begin{figure}[!b]
  \begin{center}
    \includegraphics[width=17cm,height=7cm,clip]{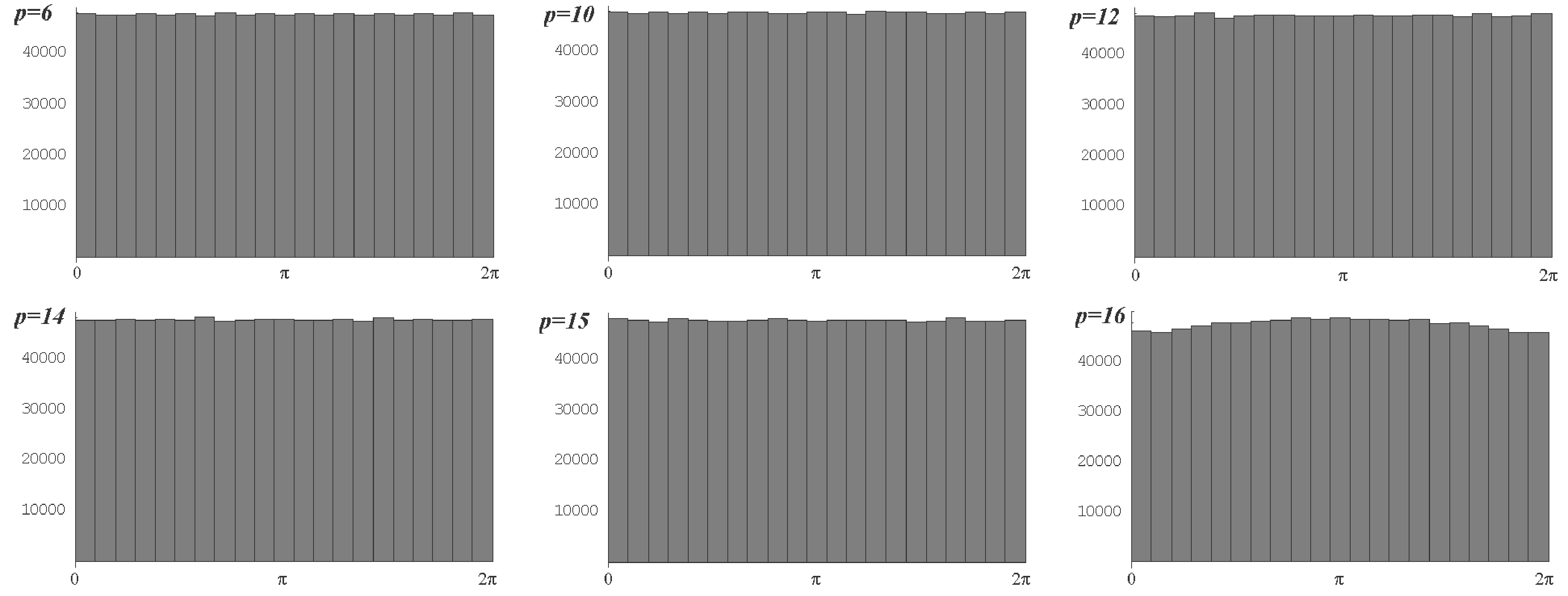}
  \end{center}
  \caption{\footnotesize The histograms(divided $21$-th) of distributions for 
           $p=6,10,12,14,15$ and $16$ beginning with $j=10^6$ up to $2\times 10^6$.}
\end{figure}

Figure 3 is the histograms of distributions for $p_k=2,3,5,7,11$ and $13$. 
In contrast to these histograms, the histograms of distributions 
in case that we put composite numbers($=6,10,12,14,15$ and $16$) into $p_k$, are also printed in Figure 4. 
In the cases for the power of one prime like $p_k=16$, a shape of the peak around $\pi$ slightly remaines 
in the histogram, whereas the shape of the tales near $0$ or $2\pi$ would be convex downwards. 

A nature of primes is also found in a distribution for the interval of succeeding primes, 
\begin{equation}
  \frac{p_{k+1}}{\log p_{k+1}}-\frac{p_k}{\log p_k},
\label{e601}
\end{equation}
where the logarithm terms exist in order to normalize to one. 
We present the histogram for $10^6$ primes beginning with $k=10^6$ in Figure 5 for example. 
The fluctuation in the histogram rather looks like an oscillation never vanish for larger number of primes 
and is deeply related to the Wilson theorem and the Hoheisel constant.

\begin{figure}[!h]
  \begin{center}
    \includegraphics[width=15cm,height=7cm,clip]{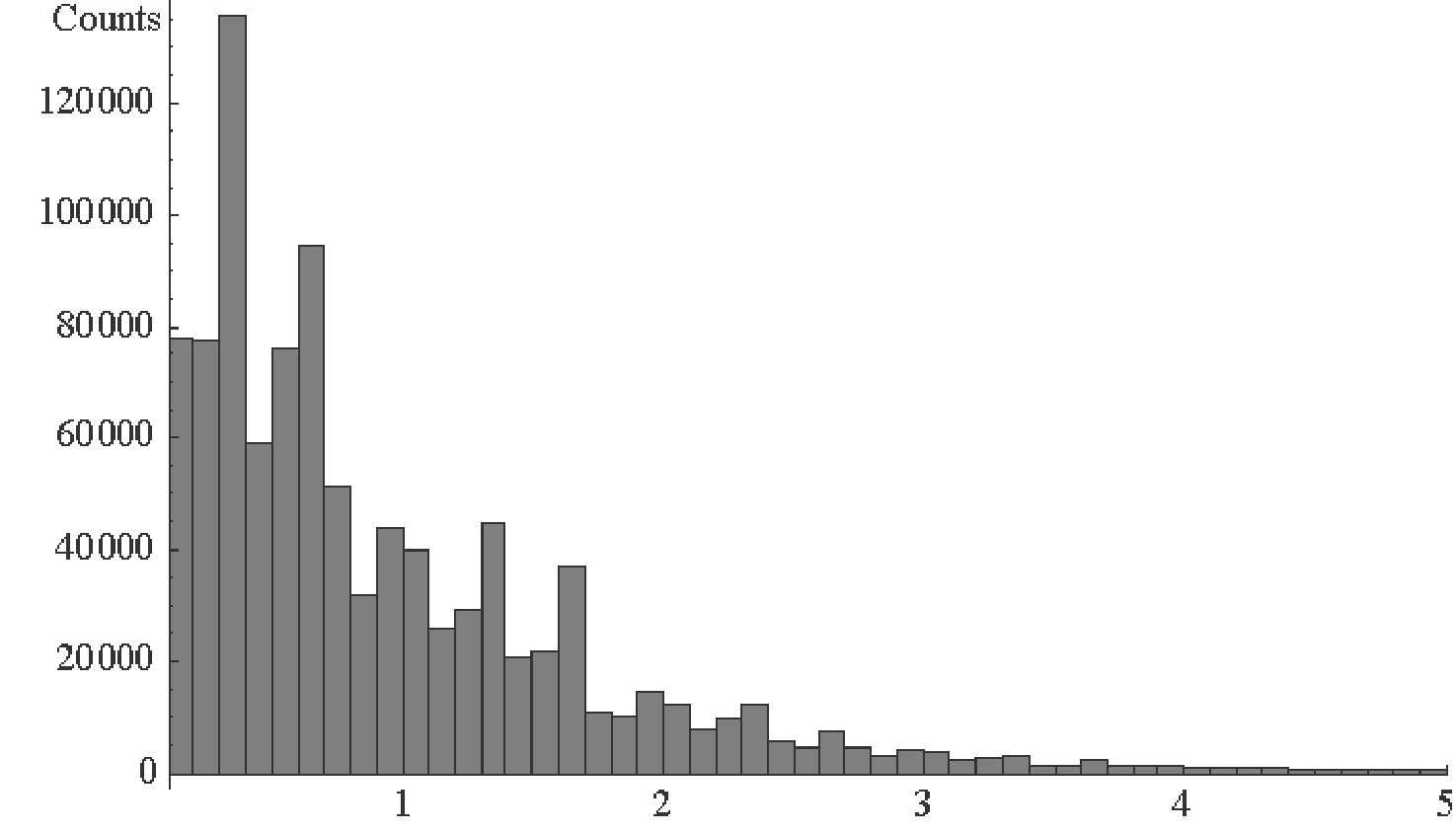}
  \end{center}
  \caption{\footnotesize The histogram(class interval $=0.1$) for the distribution of the interval of 
           $p_k/\log p_k$ for $10^6$ primes beginning with $k=10^6$.}
\end{figure}

\vskip 5mm
\section{Discussions and remarks}
\hspace{\parindent}

We discuss the equations which yield the primes and the zeros of the zeta functions 
in this section.
We normalize the product of $\lambda_j$ and $\log p_k$ introducing 
new notations $\mu_j$ and $\nu_k$ as
\begin{equation}
  \mu_j=\lambda_j,\quad
  \nu_k=\frac{\log p_k}{2\pi}, 
\label{e601}
\end{equation}
and the $k$-direction$(k=1,2,\cdots,\infty)$ average of 
$\mu_j\nu_k-\frac{1}{2}-[\mu_j\nu_k]$ will be $0$ by the distribution 
like the Sato-Tate conjucture, 
where $[\quad]$ is the Gauss symbol.
By the law of large number, we can write down 
\begin{equation}
  \sum_{k=1}^m\left(\mu_j\nu_k-\frac{1}{2}\right)-\sum_{k=1}^m[\mu_j\nu_k]
  =O\left(\frac{1}{m}\right),
\label{e603}
\end{equation}
so we get
\begin{equation}
  \mu_j=\frac{1}{\displaystyle \sum_{k=1}^m\nu_k}
        \left\{\sum_{k=1}^m[\mu_j\nu_k]+\frac{m}{2}+O\left(\frac{1}{m}\right)\right\}.
\label{e603b}
\end{equation}
We can estimate the denominator as\cite{Srivastava}
\begin{eqnarray}
  \log p_k&=&\log(k\log k+O(\log k))
          =\log k+O(\log\log k),\\
\label{e604}
  \nu_k&=&\frac{1}{2\pi}\log p_k
       =\frac{1}{2\pi}(\log k+O(\log\log k)),\\
\label{e605}
  \sum_{k=1}^m\nu_k&=&\sum_{k=1}^m\frac{1}{2\pi}\log p_k\nonumber\\
       &\simeq&\frac{1}{2\pi}\int_1^m(\log x+O(\log\log x))dx\nonumber\\
       &\simeq&\frac{1}{2\pi}m(\log m-1)+O((\log\log m)\log m).
\label{e606}
\end{eqnarray}
After all, we write a following approximate relation for any $\mu_j$, 
\begin{eqnarray}
  \mu_j&=&\frac{\displaystyle \sum_{k=1}^m[\mu_j\nu_k]+\frac{m}{2}+O\left(\frac{1}{m}\right)}
            {\frac{1}{2\pi}m(\log m-1)+O((\log\log m)\log m)}\nonumber\\
       &=&\frac{\displaystyle 2\pi\sum_{k=1}^m[\mu_j\nu_k]}{m(\log m-1)}
       +\frac{\pi}{\log m-1}
       +O\left(\frac{1}{m^2\log m}\right),
\label{e607}
\end{eqnarray}
using $\lambda_j=\mu_j,\displaystyle\nu_k=\frac{\log p_k}{2\pi}$,
we write the relation for any $\lambda_j$
\begin{equation}
  \lambda_j=\frac{\displaystyle 2\pi\sum_{k=1}^m\left[\frac{\lambda_j\log p_k}{2\pi}\right]}
                {m(\log m-1)}
          +\frac{\pi}{\log m-1}
          +O\left(\frac{1}{m^2\log m}\right).
\label{e608}
\end{equation}

In the similar way, we take the $j$-direction average of 
$\mu_j\nu_k-\frac{1}{2}-[\mu_j\nu_k]$, we can write down 
by a symmetric property as illustrated in Figure 3, 
we get 
\begin{equation}
  \nu_k=\frac{1}{\displaystyle\sum_{j=1}^n\mu_j}
        \left\{\frac{n}{2}+\sum_{j=1}^n[\mu_j\nu_k]+O\left(\frac{1}{n}\right)\right\}.
\label{e610b}
\end{equation}
We also estimate the denominator as 
\begin{eqnarray}
  \lambda_j&=&\frac{2\pi j}{\log j+\log 2\pi}+O\left(\frac{j}{\log^2j}\right)\\
\label{e611}
  \sum_{j=1}^n\lambda_j&=&\lambda_1
                       +2\pi\sum_{j=2}^n\frac{j}{\log j}
                       +\sum_{j=2}^nO\left(\frac{j}{\log^2j}\right)\nonumber\\
  &=&2\pi\int_2^n\frac{x}{\log x}dx
            +\int_2^nO\left(\frac{x}{\log^2x}\right)dx\nonumber\\
  &=&2\pi\left[\frac{x^2}{2\log x}\right]_2^n
            +\pi\int_2^n\frac{x}{\log^2x}dx
            +\left[O\left(\frac{j}{\log^2j}\right)\right]_2^n\nonumber\\
  &=&\frac{\pi n^2}{\log n}
            +O\left(\frac{n^2}{\log^2n}\right)=\sum_{j=1}^n\mu_j
\label{e612}
\end{eqnarray}
so we write a following approximate relation for any $\nu_k$,
\begin{eqnarray}
  \nu_k
       &=&\frac{\displaystyle\sum_{j=1}^n[\mu_j\nu_k]+\frac{n}{2}+O\left(\frac{1}{n}\right)}
               {\displaystyle\frac{\pi n^2}{\log n}
                             +O\left(\frac{n}{\log^2n}\right)}\nonumber\\
       &=&\frac{1}{2\pi}\left(\frac{\displaystyle 2\log n\sum_{j=1}^n[\mu_j\nu_k]}{n^2}
                              +\frac{\log n}{n}\right)
         +O\left(\frac{\log n}{n^3}\right).
\label{e614}
\end{eqnarray}
Finally we write down the relation for any $p_k$
\begin{eqnarray}
  \log p_k&=&2\pi\nu_k
          =\frac{\displaystyle 2\log n\sum_{j=1}^n\left[\frac{\lambda_j\log p_k}{2\pi}\right]}
                            {n^2}
                       +\frac{\log n}{n}
          +O\left(\frac{\log n}{n^3}\right)\\
\label{e615}
  p_k&=&\exp\left\{\frac{2\log n\sum_{j=1}^n\left[\frac{\lambda_j\log p_k}{2\pi}
                            \right]}{n^2}
                       +\frac{\log n}{n}\right\}\cdot
        \exp\left\{O\left(\frac{\log n}{n^3}\right)\right\}\nonumber\\
     &=&n^{\left(\frac{2\sum_{j=1}^n\left[\lambda_j\log p_k/(2\pi)\right]}{n^2}
                 +\frac{1}{n}\right)}\cdot
        n^{O\left(\frac{1}{n^3}\right)}
\label{e616}
\end{eqnarray}
Equations (\ref{e608}) and (\ref{e616}) are a set of equations which gives prime numbers 
and zeros of the Riemann zeta function.

\vskip 5mm

\end{document}